\documentclass[iop,numberedappendix]{emulateapj}

\newcommand{ \rmxaa}{The Revista Mexicana de Astronomía y Astrofísica}
\newcommand{\kms}{\mbox{\,km$\,$s$^{-1}$}}

\newcommand{\hfortytwo}{H42$\alpha$\,}

\newcommand{\lda}{\bf COMMENTS from LDA: }

 \slugcomment{1429150517}

\begin{document}

   \title{A Large-scale Map of Millimeter Hydrogen Recombination Lines around a Super Star Cluster
   }

   


\author{Q. Nguy$\tilde{\hat{\rm e}}$n-Lu{\hskip-0.65mm\small'{}\hskip-0.5mm}o{\hskip-0.65mm\small'{}\hskip-0.5mm}ng\altaffilmark{1,2}}
\affil{Korea Astronomy and Space Science Institute, 776 Daedeok daero, Yuseoung, Daejeon 34055, Republic of Korea}
\affil{NAOJ Chile Observatory, National Astronomical Observatory of Japan, 2-21-1 Osawa, Mitaka, Tokyo 181-8588, Japan}

\author{L.D. Anderson}
\affil{Department of Physics and Astronomy, West Virginia University, Morgantown, WV 26506, USA}
\affil{Adjunct Astronomer at the National Radio Astronomy Observatory, P.O. Box 2, Green Bank, WV 24944, USA)} 
\affil{Center for Gravitational Waves and Cosmology, West Virginia University, Chestnut Ridge Research Building,
Morgantown, WV 26505, USA}

\author{F. Motte}
\affil{Institut de Planétologie et d'Astrophysique de Grenoble, Univ. Grenoble Alpes - CNRS-INSU, BP 53, 38041, Grenoble Cedex 9, France}

\author{K.T. Kim}
\affil{Korea Astronomy and Space Science Institute, 776 Daedeok daero, Yuseoung, Daejeon 34055, Republic of Korea}
\author{P. Schilke, P. Carlhoff, N.  Schneider}
\affil{I. Physik. Institut, University of Cologne, 50937 Cologne, Germany}

\author{H. Beuther, S. Bihr, M. Rugel, J. Soler, Y. Wang}
\affil{Max Planck Institute for Astronomy, Konigstuhl 17, 69117 Heidelberg, Germany}

\author{P. Didelon}
\affil{Laboratoire AIM Paris-Saclay, CEA/IRFU - CNRS/INSU - Universit\'e Paris Diderot, Service d'Astrophysique, B\^at. 709, CEA-Saclay, F-91191, Gif-sur-Yvette Cedex, France}
\author{C. Kramer}
\affil{Instituto Radioastronomía Milimétrica (IRAM), Av. Divina Pastora 7, Nucleo Central, 18012, Granada, Spain}
\author{F. Louvet, L. Bronfman}
\affil{Departamento de Astronom\'ia, Universidad de Chile, Santiago, Chile}

\author{K. M. Menten}
\affil{Max-Planck-Institut für Radioastronomie, Auf dem Hügel 69, 53121, Bonn, Germany}

\and

\author{C. M. Walmsley\altaffilmark{3}}


\altaffiltext{1}{EACOA Fellow at KASI and NAOJ}
\altaffiltext{2}{quangnguyenluong@kasi.re.kr}
\altaffiltext{3}{This work is dedicated to Walmsley C. M. (1941--2017), a member of W43-HERO, who pioneered in using recombinations lines as diagnostic of the ionized gas component of the interstellar medium  \citep{hoangbinh74, walmsley75,churchwell75, mebold76, weisheit77, shaver78,pankonin79, walmsley81,walmsley82,walmsley90,gordon90,sorochenko91,natta94,smirnov95, wyrowski97}.}

\begin{abstract}
We report the first map of large-scale (10 pc$^{2}$) emission of millimeter hydrogen recombination lines (mm-RRLs) toward the giant H{\scriptsize II} region around the W43 super star cluster.  Our mm-RRL observations come from the IRAM 30m telescope, and are analyzed together with radio continuum and cm-RRLs data from the Karl G. Jansky Very Large Array (VLA) and HCO$^{+}$ 1--0 line emission data from the IRAM 30m. The mm-RRLs reveal an expanding wind-blown ionized gas bubble with an electron density $n_{\rm e}\sim70-1500$\,cm$^{-3}$ driven by the WR/OB cluster, which produces a total Lyman $\alpha$ photon rate of $1.5\times1-^{50}$ per second as calculated from the radio continuum emission. This bubble is interacting with the dense gas in the W43-Main ridge. Combining the high spectral and angular resolution cubes of mm-RRL and cm-RRL, we derive the 2-dimensional relative distributions of dynamical and pressure broadening of the ionized gas and find that the RRL line widths are broadened mostly by pressure broadening (4--55\,\kms) inside the cavity and mostly by dynamical broadening (8--36\,\kms) near the bubble edge. Ionized gas clumps found at the edge of the bubble may suggest that large-scale ionized gas motion causes swept-up gas in this region and triggers the formation of Ultra Compact H{\scriptsize II} regions and the massive young stellar object near the periphery of the giant wind-blown bubble. 
\end{abstract}



\section{Introduction} \label{sec:intro}
Massive stars are not formed alone but in clusters, either sequentially or simultaneously (see review in \citealt{tan14}). Young embedded massive protoclusters are often found at the junctions or hubs of a filamentary network where matter accretion occurs continuously \citep{hill11,schneider12}. Due to the large amount of energy 
that massive stars inject into their local environments, as soon as reaching the main-sequence they will develop an H{\scriptsize II} region around them that contains mostly ionized hydrogen gas \citep[e.g.][]{churchwell02}.  

Ionized gas can be probed using observations of radio continuum and recombination lines \citep{natta94,smirnov95}.  For star-forming regions that suffer from high extinction, radio recombination lines (RRL, \citealt{walmsley81}), including those at millimeter wavelengths (mm-RRL, \citealt{gordon90,walmsley90}), can be used instead of those at shorter wavelengths. The mm-RRLs have two advantages over cm-regime RRLs. First, they are thermally broadened by dynamical motions, allowing the determination of ionized gas kinematics \citep{keto08}. Second, mm-RRLs are stronger than RRLs, though high electron densities ($n_{\rm e}\sim10^{7-8}$\,cm$^{-3}$) are required to amplify them \citep{walmsley90,peters12}. To date, most of the observations of mm-RRLs have used only single pointings or small interferometric maps toward strong, compact sources  \citep{keto08,galvan-madrid12,kim17}. mm-RRLs were also detected in the 3mm waveband surveys of the Central Molecular Zone of the Milky Way but its spatial extent is compact and confined toward SgrB2 \citep{jones12}. 

The giant H{\scriptsize II} region W43 is powered by the W43 super star cluster (SSC), a prototypical example of a SSC \citep{fukui14}. In the W43-SSC, more than 50 OB stars contribute to a strong infrared continuum luminosity of $\sim3.5\times^6 L_{\sun}$ \citep{lester85,blum99}. 
Radio continuum observations reveal an extended bubble structure \citep{balser01}, later confirmed by the HI/OH/Recombination line survey of the inner Milky Way (THOR) \citep{beuther16,bihr16}.  
To the west of the SSC there exist the younger embedded massive protocluster W43-MM1 and W43-MM2, part of the W43-Main dense clouds, which are in the process of forming massive stars efficiently \citep{motte03,louvet14}. 
Active massive star formation sites such as High-Mass Starless Clumps (HMSLCs) \citep{beuther12} are also found further from the W43-Main dense clouds, in the bubble and chimney toward the west of W43-MM1 and the W43-MM2 ridge \citep{bally10}. On larger scales, the W43-Main cloud is embedded inside the active W43 massive cloud complex \citep{nguyenluong11}. The intense ministarburst activity of W43 is continuously fueled by material, owing to its position at the meeting point of the Scutum-Centaurus (or Scutum-Crux) Galactic arm and the bar \citep{nguyenluong11,carlhoff13,motte14}.  
Recently, spatially extended low-velocity shocks traced by SiO emission, possibly arising from cloud-cloud collisions, were discovered in the W43-MM1 and W43-MM2 ridges \citep{nguyenluong13,louvet16}. These marked the boundaries between ionized and dense molecular gas. The morphology of the ionized and molecular gas is suggestive of an interaction \citep{motte03}. 

In this Letter, we report the first-ever large-scale high angular and spectral resolution mm-RRL map and provide the first direct kinematic evidence of the interaction between ionized and molecular gas in the region.

\section{Observational data}
\label{sect:observation}

\begin{figure*}[htbp!]
\centering
$\begin{array}{ccc}
\includegraphics[angle=0,width=5.2cm]{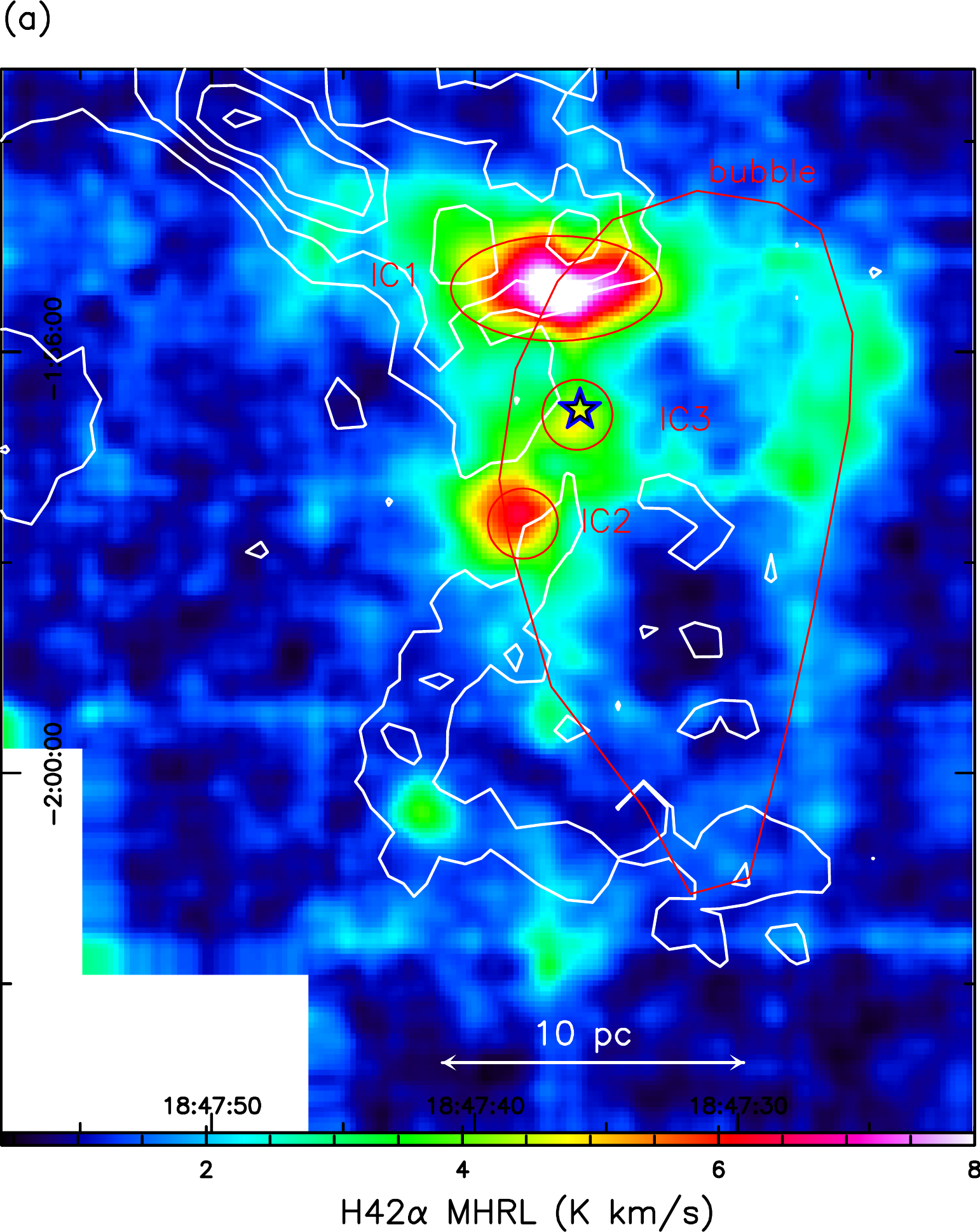}& 
\includegraphics[angle=0,width=5.2cm]{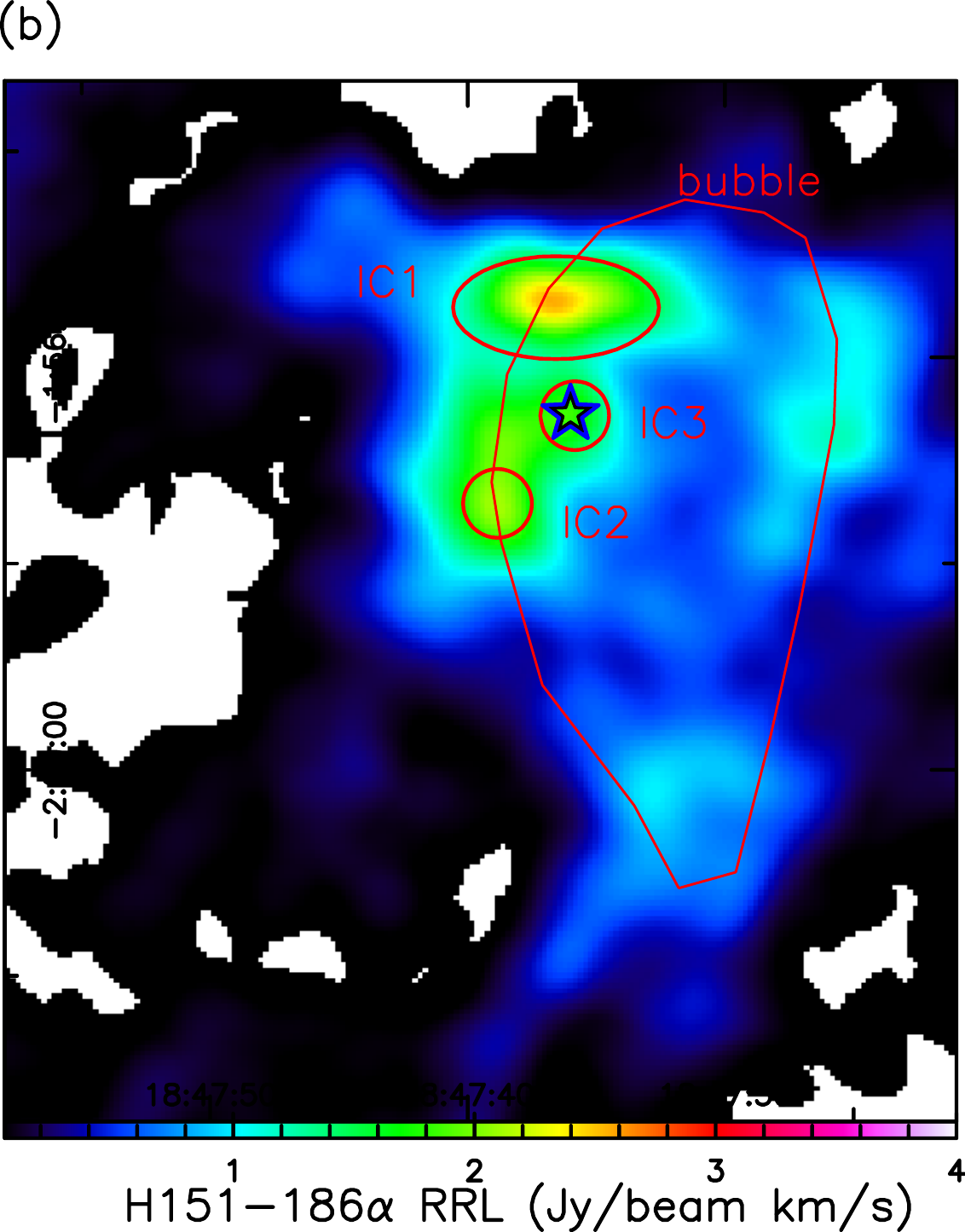}&
\includegraphics[angle=0,width=5.4cm]{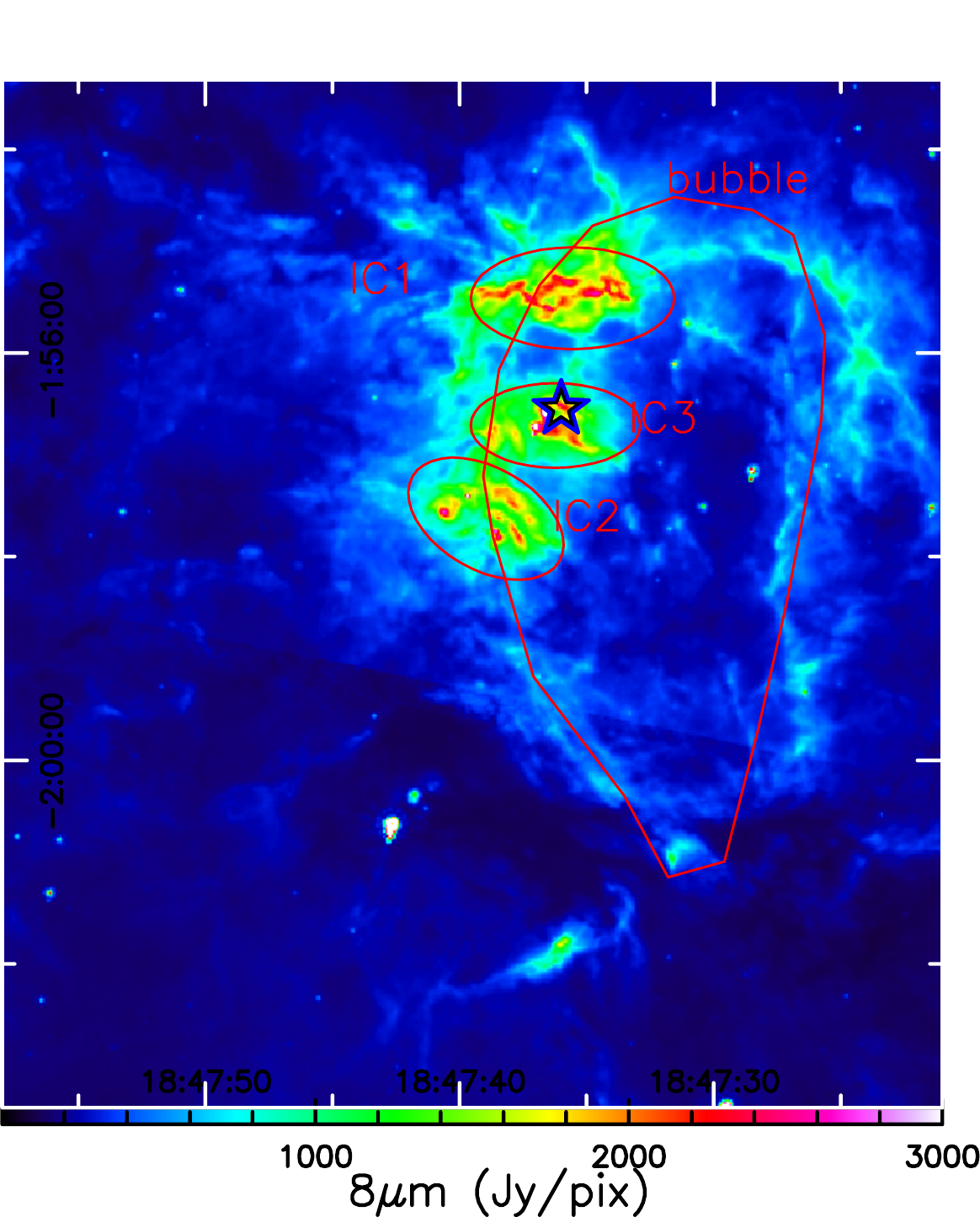} 

\end{array}$
\caption{{\bf (a)}: Integrated intensity map of H42$\alpha$ (color), integrated over 70--114\,\kms, and HCO$^{+}$ 1--0 (contours), in steps of 10, 15, 20, and 25 K$\kms$; {\bf (b)}: Integrated intensity map of cm-RRLs from THOR, also integrated over 70--114\,\kms; and
 {\bf (c)}: The 8$\micron$ {\it Spitzer} image,
On all panels, red polygon denotes the position of the ionized gas bubble and red ellipses indicate the position of the ionized gas clumps.  The blue star marks the position of the W43 SSC \citep{blum99}. 
}
\label{fig:recombinationlines}
\end{figure*}

This Letter makes use of hydrogen recombination lines H$41\alpha$  ($\nu_{\rm rest}$ = 92.034 GHz) and H$42\alpha$ ($\nu_{\rm rest}$ = 85.688 GHz),  as well as HCO$^{+}$ 1-0 ($\nu_{\rm rest}$ = 89.188 GHz) datacubes from the W43-HERO IRAM 30~m Large Program ``Origins of molecular clouds and star formation in W43''(PIs: Fr\'ed\'erique Motte and Peter Schilke, \citealt{nguyenluong13,carlhoff13}). W43-HERO
 conducted an 8~GHz bandwidth line mapping survey at the wavelengths of 3~mm (85--93~GHz) toward W43-Main using the Eight MIxer Receiver (EMIR) and the large bandwidth (8~GHz) Fast Fourier Transform Spectrometer (FTS) backend at 200 kHz resolution\footnote{\tt The main repository of the data is hosted by IRAM at http://www.iram.fr/ILPA/LP004/}. 
The reduced spectra are combined into
gridded data cubes with a $10\arcsec$ Gaussian kernel. The effective angular resolution, spectral resolution, and rms noise of these three lines are 28\arcsec, $0.67\,\kms$, and 17--50 mK per channel. Further explanation of the observation data reduction can be found in \cite{nguyenluong13}.

Additionally, we use the radio 1.436 GHz continuum map and the stacked RRL cube from 19 hydrogen recombination lines (151$\alpha$ to 186$\alpha$) emitting in the frequency range 1.0--1.9 GHz. These data are from ``The HI/OH/Recombination line survey of the inner Milky Way (THOR)'' large program with the Karl G. Jansky Very Large Array (VLA) \citep{beuther16}. 
The final data has an angular resolution of $25\arcsec$ and a continuum sensitivity of 0.6 mJy/beam. The RRL cube is smoothed to an angular resolution of 40\arcsec\, and a spectral resolution of 10\,\kms, and it has an rms noise of of 3 mJy/beam per channel. Both THOR data sets lack short-spacing information. Further explanation of the observations, data reductions, and line stacking methods can be found in \cite{beuther16}. The W43-Main ionized gas nebula is also named as G30.782-0.027 in THOR continuum source catalog \citep{bihr16}. 

\section{The large scale ionized gas bubble around the W43-Main starburst cluster}
\label{sect:results}
Because the two mm-RRL lines in our study are at adjacent principle quantum levels, 
they trace emission from the same ionized gas components.  Hence, we use only the \hfortytwo line for our analysis.

\subsection{The 10 parsec-scale mm-RRL map}
We show the first example of an extended (10 pc) mm-RRL map surrounding an SSC in the integrated intensity map of \hfortytwo emission in Figure\,\ref{fig:recombinationlines}a. The emission is integrated from 70 to 114\kms, which is $\sim10\kms$ larger in the blue-shifted part and 4\,$\kms$ larger in the red-shifted part when compared to the main velocity range of molecular gas in W43-Main \citep{nguyenluong11}.

The morphology of mm-RRL emission is different from that of dense gas tracers. The notable feature of the integrated mm-RRL map is the boxy bubble with an inner cavity dominating at the right edge of the Z-shape filaments of W43-Main ridge as seen in HCO$^{+}$ 1--0 (see Figure\,\ref{fig:recombinationlines}a-b). The bubble is well-defined in mm-RRLs, cm-RRLs, and $8\,\micron$ emission. In the North-South direction, the mm-RRL emission extends $\sim15$pc, while it is about 10pc in the East-West direction (see Figure\,\ref{fig:recombinationlines}). The strong 8\,$\micron$ emission shows the effect of radiation that heats the dust and PAH particles. Given the nearly symmetric morphology of the bubble, the SSC is the ionizing source responsible for
producing the bubble. 

In comparison with mm-RRLs, the cm-RRL integrated emission as well as radio continuum emission show a similar morphology but it, although the cm-RRLs do have more diffuse emission surrounding the boxy shape around the W43-SSC (see Figure\,\ref{fig:recombinationlines}b).  
The maximum brightness temperature of the radio continuum emission is $T_{\rm cont} \sim 2200$ K and the average spectral index is 0.31. These show that the ionized gas in W43-Main is dominated by free-free emission \citep{bihr16}. In addition to the large-scale bubble, we notice that several bright ionized gas clumps (e.g., IC1, IC2, and IC3) exist at the periphery of the bubble. This is especially noticeable to the west of W43-SSC where large-scale ionized gas and dense gas interact.  The bubble is surrounded by a thin filamentary shell and the ionized clumps are resolved to spider-web filamentary structures in IC1, and bow-shock structure in IC2 and IC3. The 8\,$\micron$ image in Figure\,\ref{fig:recombinationlines}c shows hot dust that coincides with the location of three ionized clumps and the ionized bubble. 
 
  \begin{figure*}[tph!]
\centering
$\begin{array}{ccccc}
\hspace{-0.4cm}
\includegraphics[angle=0,width=3.5cm]{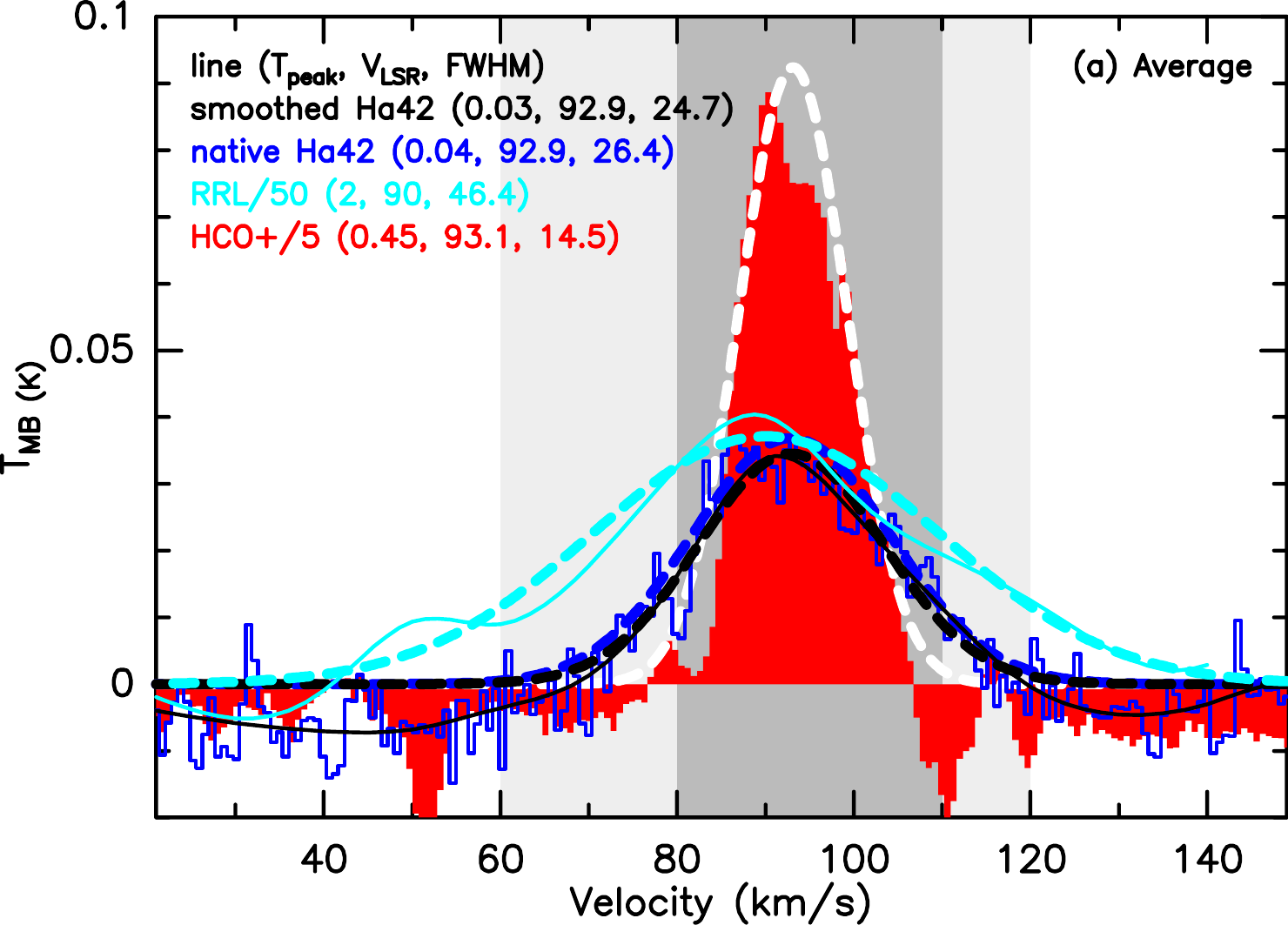} &
\includegraphics[angle=0,width=3.5cm]{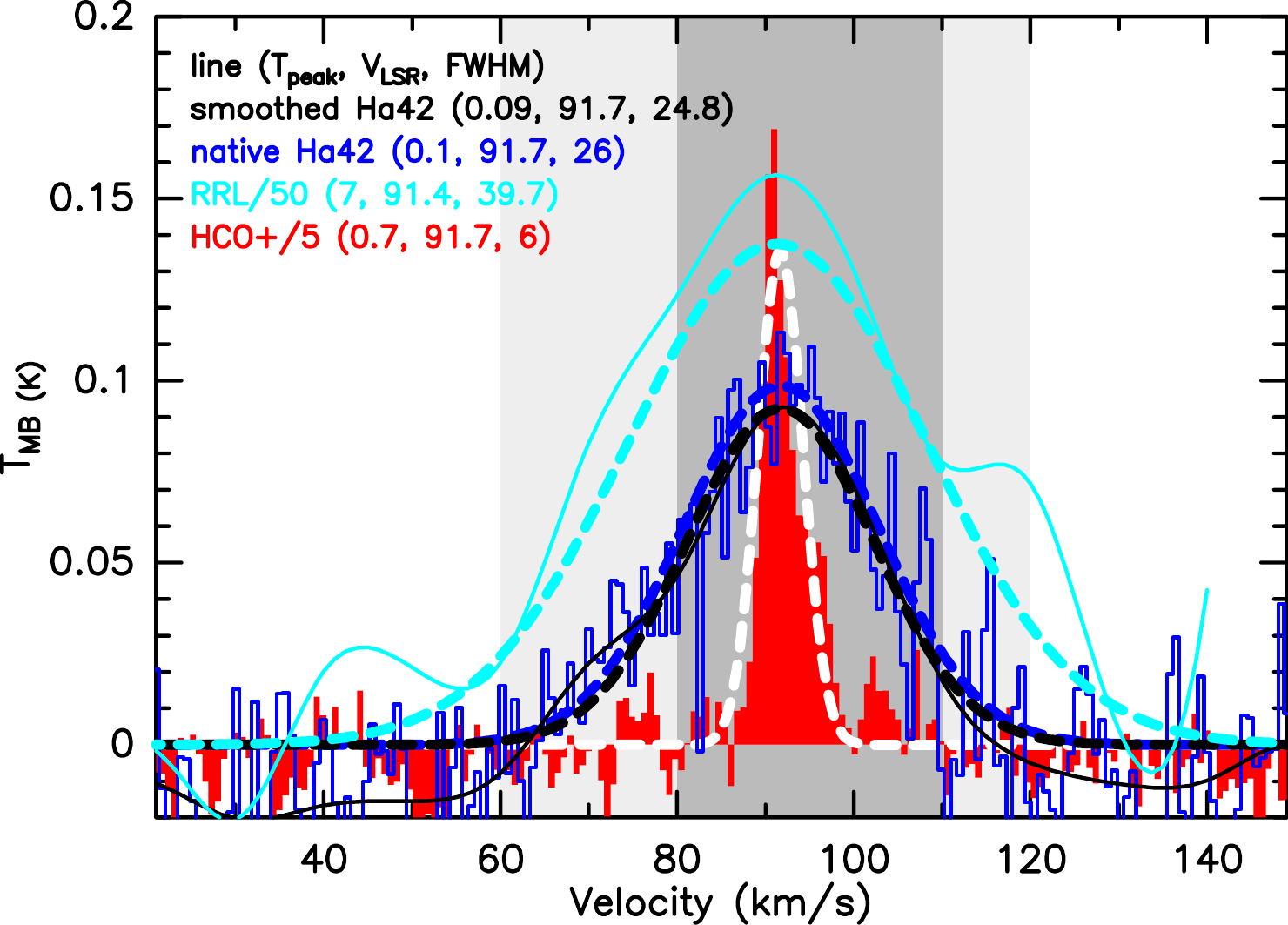} &
\includegraphics[angle=0,width=3.5cm]{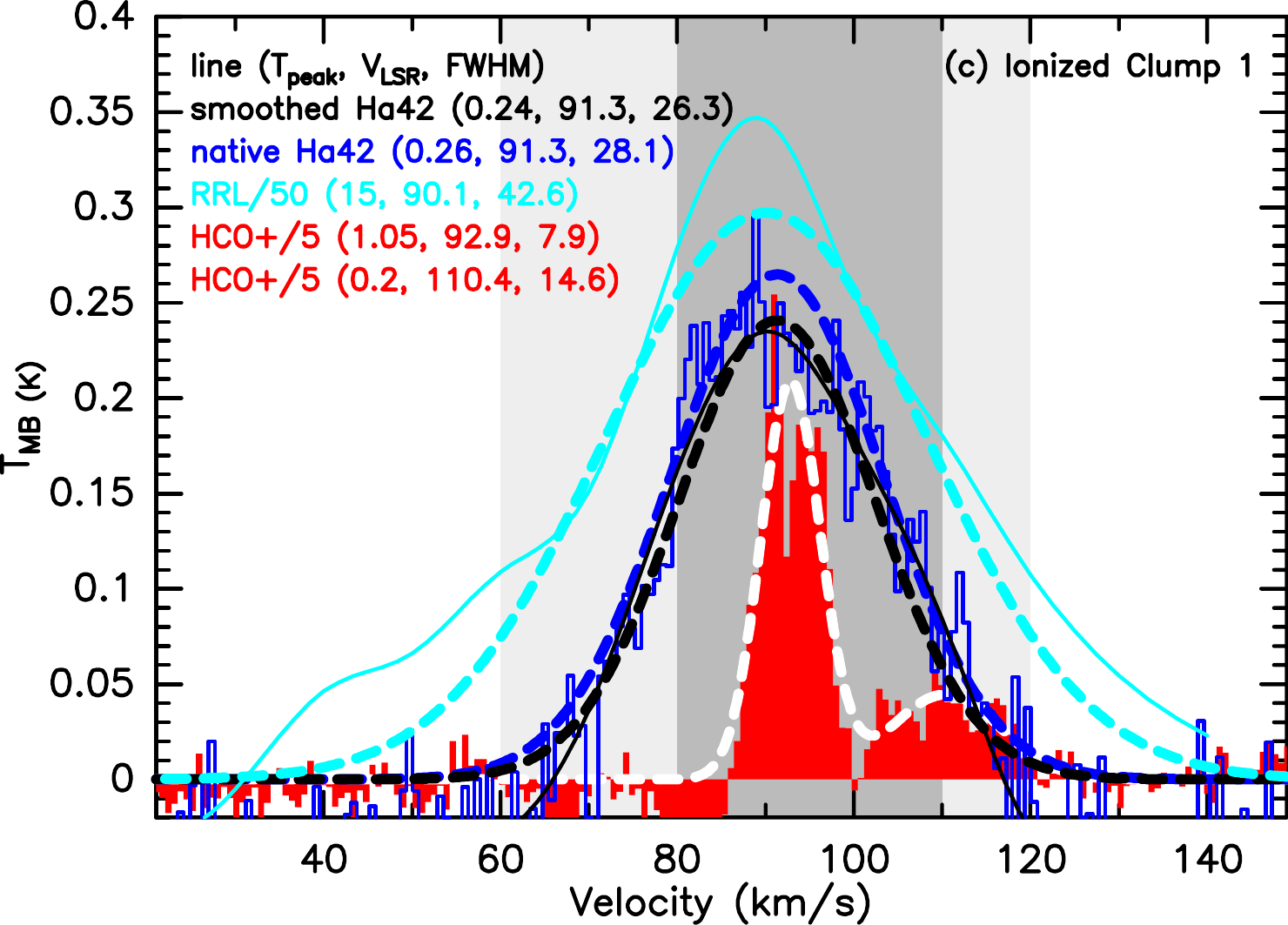}  &
\includegraphics[angle=0,width=3.5cm]{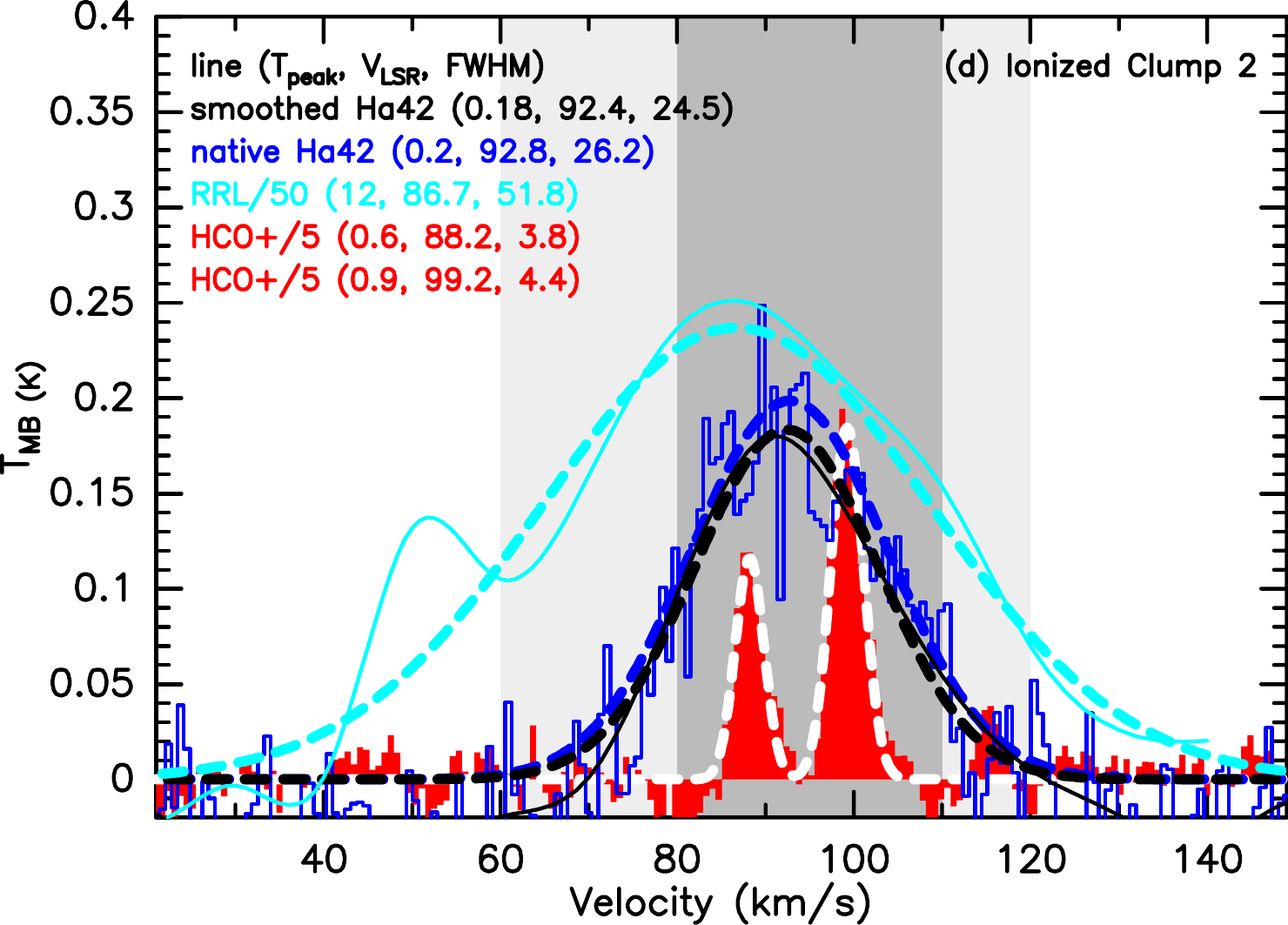}  &
\includegraphics[angle=0,width=3.5cm]{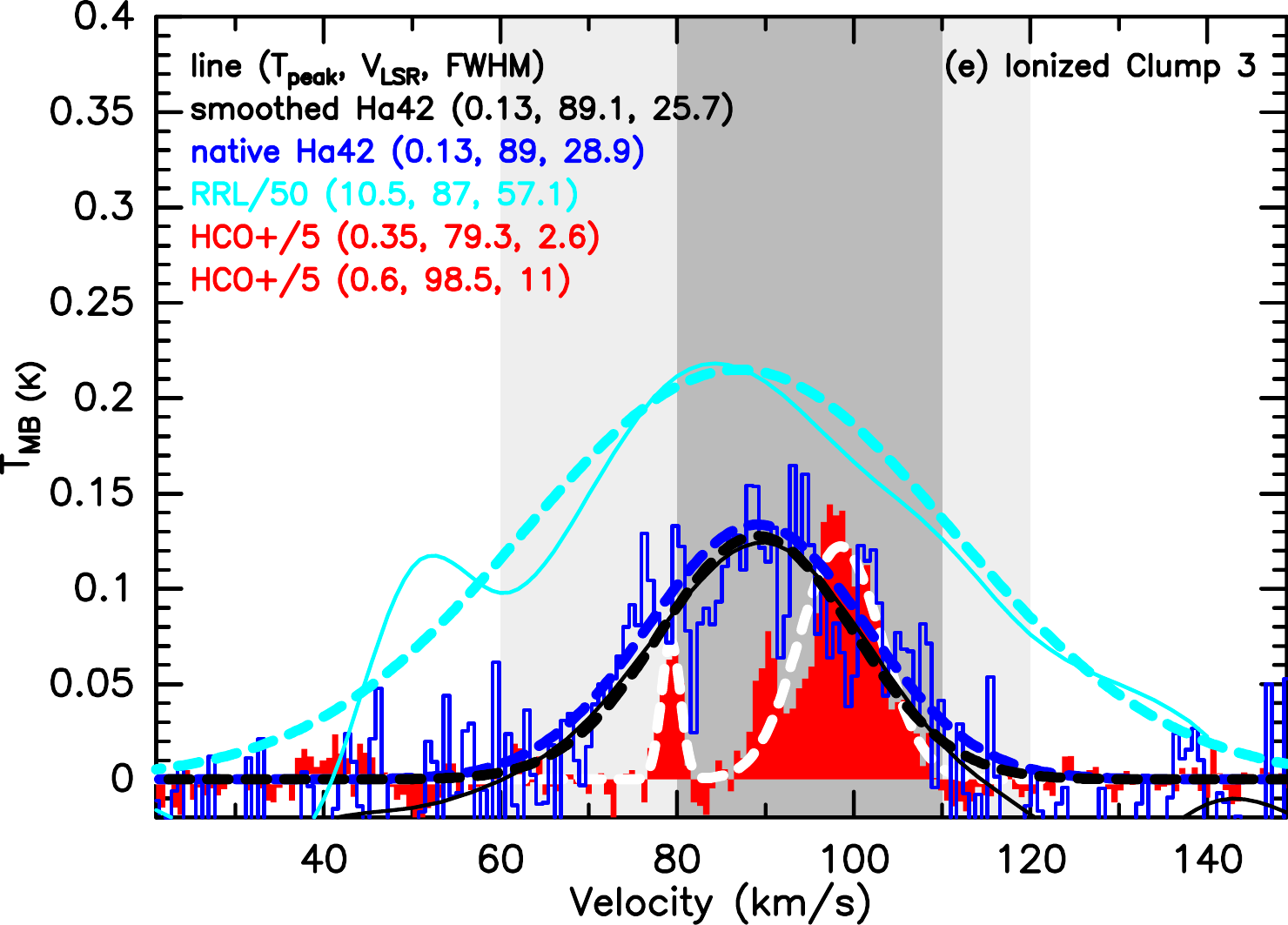} 
\end{array}$
\caption{Spectra of H42$\alpha$ at its native resolution (dark blue),  H42$\alpha$ smoothed to similar angular and spectral resolution as the cm-RRLs (black), cm-RRLs (light blue), HCO$^{+}$ 1--0 (red)  and their Gaussian fits (same color dashed curves) averaged over the entire map {\bf (a)}, of the bubble ionized gas {\bf (b)}, of IC1 {\bf (c)}, of IC2 {\bf (d)}, and of IC3 {\bf (e)}. The darker-shaded area is the main velocity range of dense gas in W43-Main and the bright-shaded area is the main velocity range of the diffuse cloud traced by CO gas \citep{nguyenluong11}. Results from Gaussian fits ($T_{\rm peak}$, $V_{\rm LSR}$, FWHM) follow the line names shown on each plot. }
\label{fig:spectra}
\end{figure*}

\subsection{Expanding ionized gas}
\label{sect:expanding}
The high spectral resolution map of the large-scale mm-RRL emission around W43-SSC also shows that the ionized gas created by the giant H{\scriptsize II} region is indeed interacting with the dense molecular gas in the W43-Main ridge. 
In Figure\,\ref{fig:spectra}, we plot the cm-RRL, mm-RRL, and HCO$^{+}$ spectra and their Gaussian fits at several locations: the whole averaged map, the bubble, and the ionized clumps (IC1, IC2, and IC3). The interweaving of dense, neutral molecular gas traced by HCO$^{+}$ and ionized gas traced by mm-RRLs in W43-Main is visible by the similar velocity ranges of the average spectra around 80-110\,\kms. The mm-RRL peaks is within the main velocity range 80--115\,\kms, of W43-Main (shaded area). However, the ionized gas and neutral molecular gas are not completely overlapped as shown by velocity shifts in their line-of-sight velocities (see Figure\,\ref{fig:spectra}). The mm-RRL width is larger than that of HCO$^+$ in all positions, and is stronger than HCO$^+$ line in the red-shifted parts. In IC1 and IC2, the mm-RRLs have broadened blue-shifted wings, showing evidence of the front, possibly outflowing gas. The FWHMs of mm-RRLs are larger than pure, thermally broadened lines by 5--20\,\kms (see Section\,\ref{section:discussion})  which provides us with further evidence of outflowing ionized gas. The FWHMs of cm-RRLs  ($\sim28\,\kms$) are broader than those of mm-RRLs ($\sim18\,\kms$), which shows pressure broadening of the RRLs lines by collisions with electrons and ions (see Section\,\ref{section:discussion}). 

\begin{figure*}[tbhp!]
\centering
$\begin{array}{c}
\hspace{-0.8cm}
\includegraphics[angle=0,height=18cm]{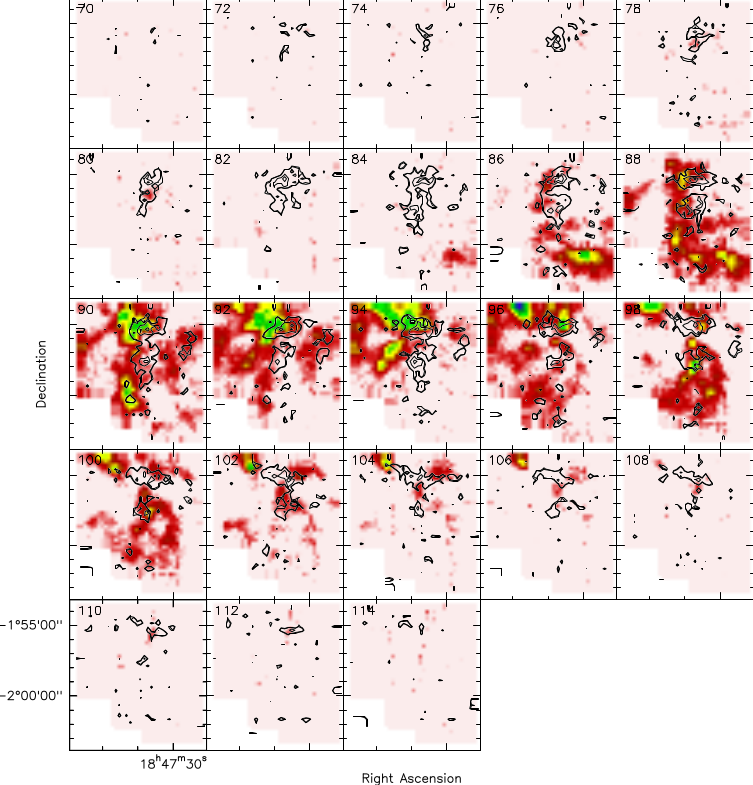}
\end{array}$
\caption{Channel maps (74--102\,\kms) of H42$\alpha$ (contour) and HCO$^{+}$ 1--0 (colours in in levels 0.08, 0.14, 0.20, 0.26 K per 2\,\kms channel) at the native resolutions, in increments of 2\,\kms. }
\label{fig:H92}
\end{figure*}

The mm-RRL channel maps (Figure\,\ref{fig:H92}) show a clear positional misalignment between ionized and neutral molecular gas traced by HCO$^{+}$. mm-RRL emission lies closer to the SSC cluster and has a sharp separation with HCO$^{+}$, especially in the 86\,\kms\, and 92\,\kms\, channels. At 94\,\kms, however, ionized gas and HCO$^{+}$ seem to merge near the location of the W43-MM1 ridge. This is also the location where extended SiO shocks from cloud-cloud-collisions were discovered \citep{nguyenluong13}. The North-South extension of the ionized gas emission is most prominent in the blue-shifted part (84-95 \kms) and surrounds the SSC. Blue-shifted emission around the 76-80\,\kms\, channels is found further from the central ionizing sources, indicating the expanding extent of the ionized gas. All channels show clear evidence of the ionized gas cavity where HCO$^{+}$ is emitting around 90--92\,\kms.
IC2 and IC3 are strong in almost all channels from 76 to 110\,\kms. They bear the general properties of  broad hydrogen recombination lines of UCH{\scriptsize II} regions caused by the outflowing gas \citep{sewilo11}.

\section{Ionized gas dynamics}
\label{section:discussion}
\begin{figure*}[htbp!]
\centering
$\begin{array}{ccccc}
\includegraphics[angle=0,width=3.2cm]{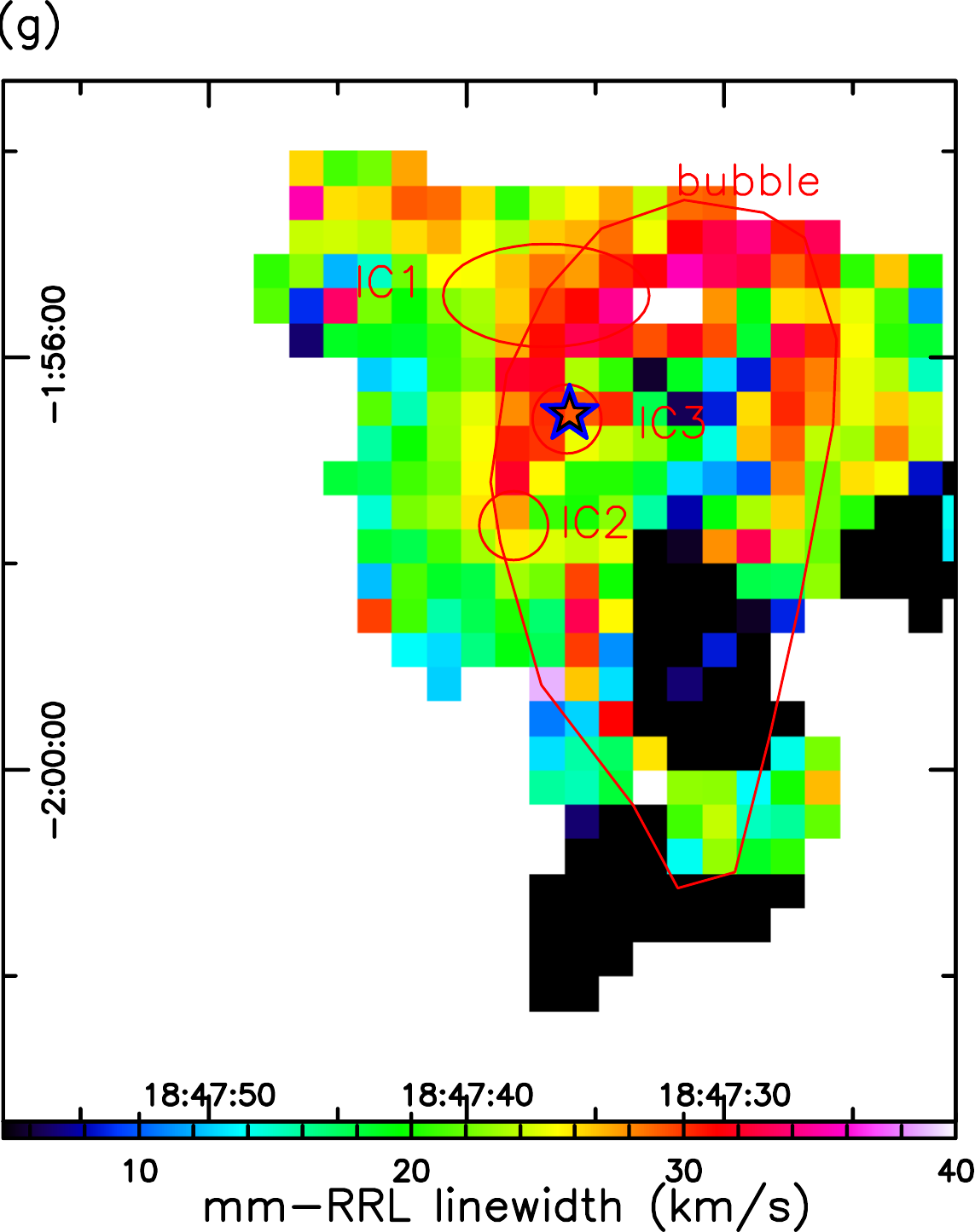} &
\includegraphics[angle=0,width=3.2cm]{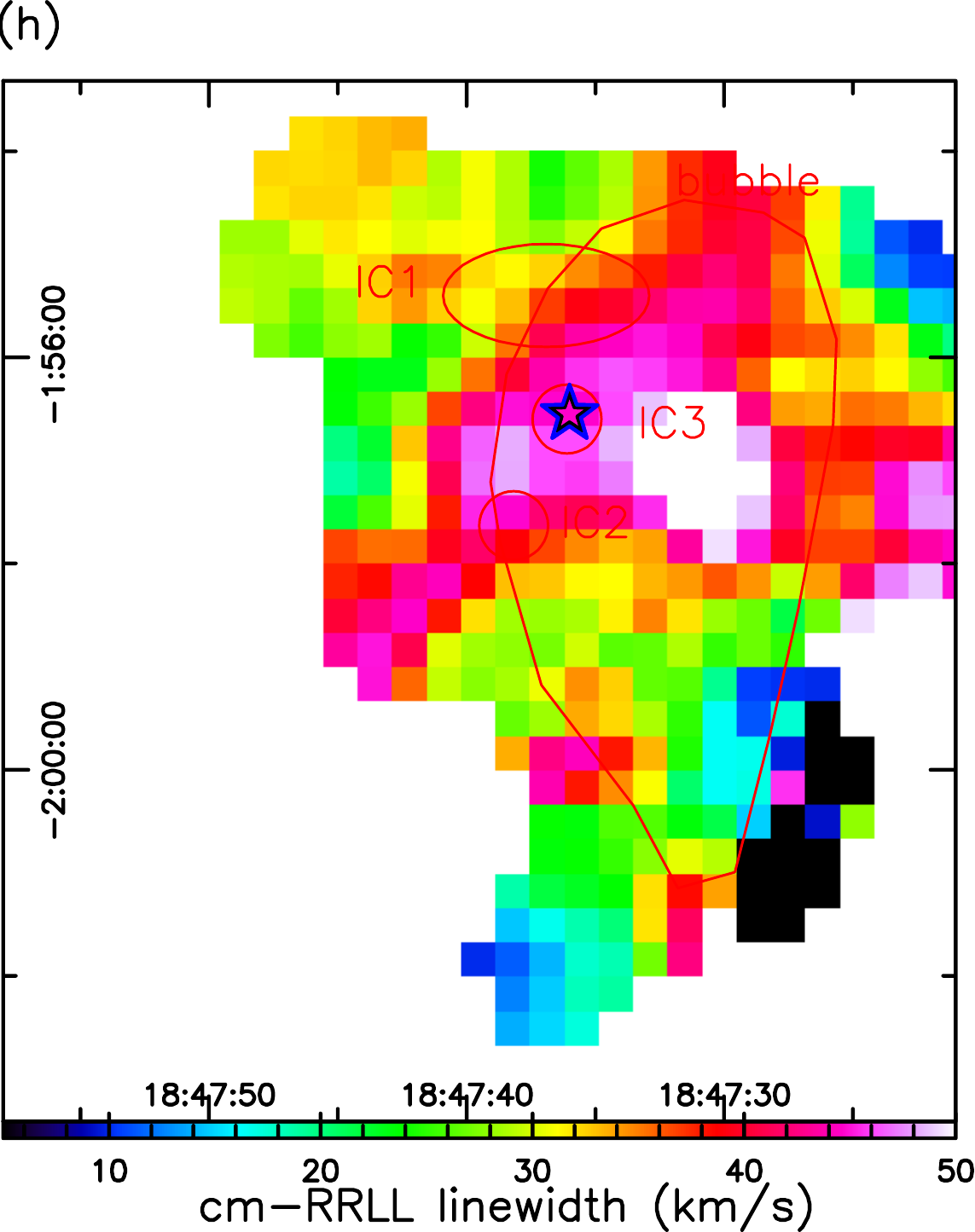} &
\includegraphics[angle=0,width=3.2cm]{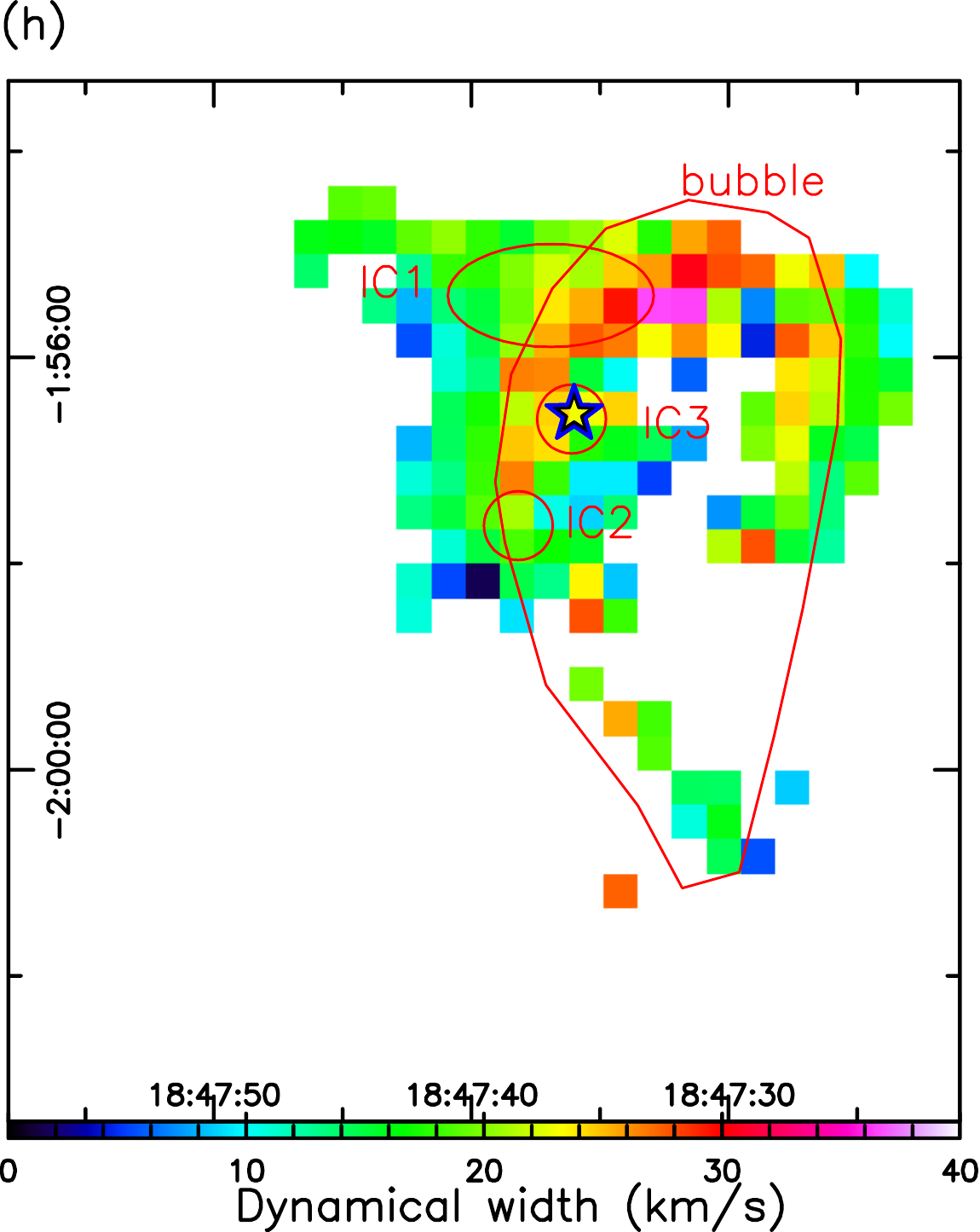}&
\includegraphics[angle=0,width=3.2cm]{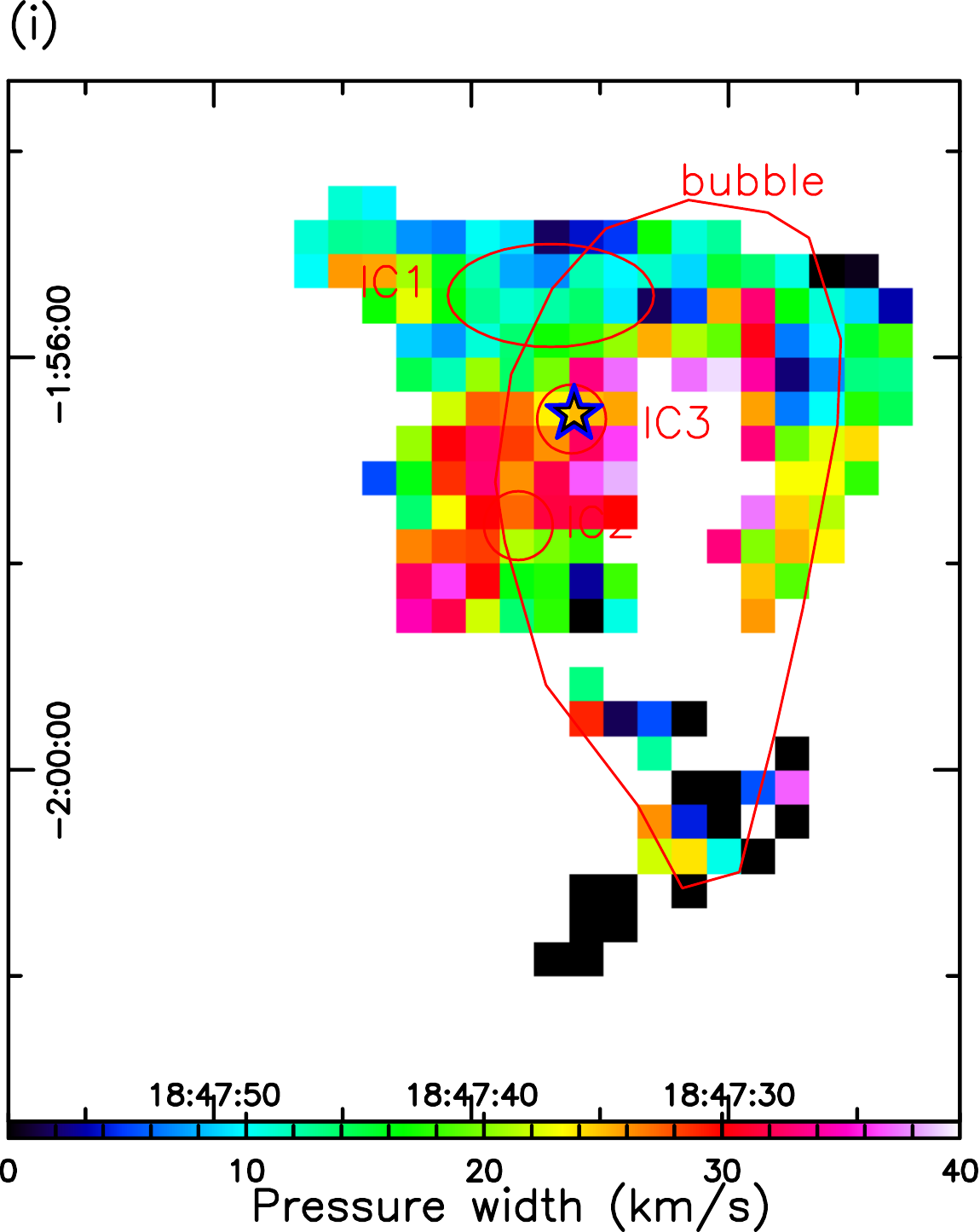}&
\includegraphics[angle=0,width=3.2cm]{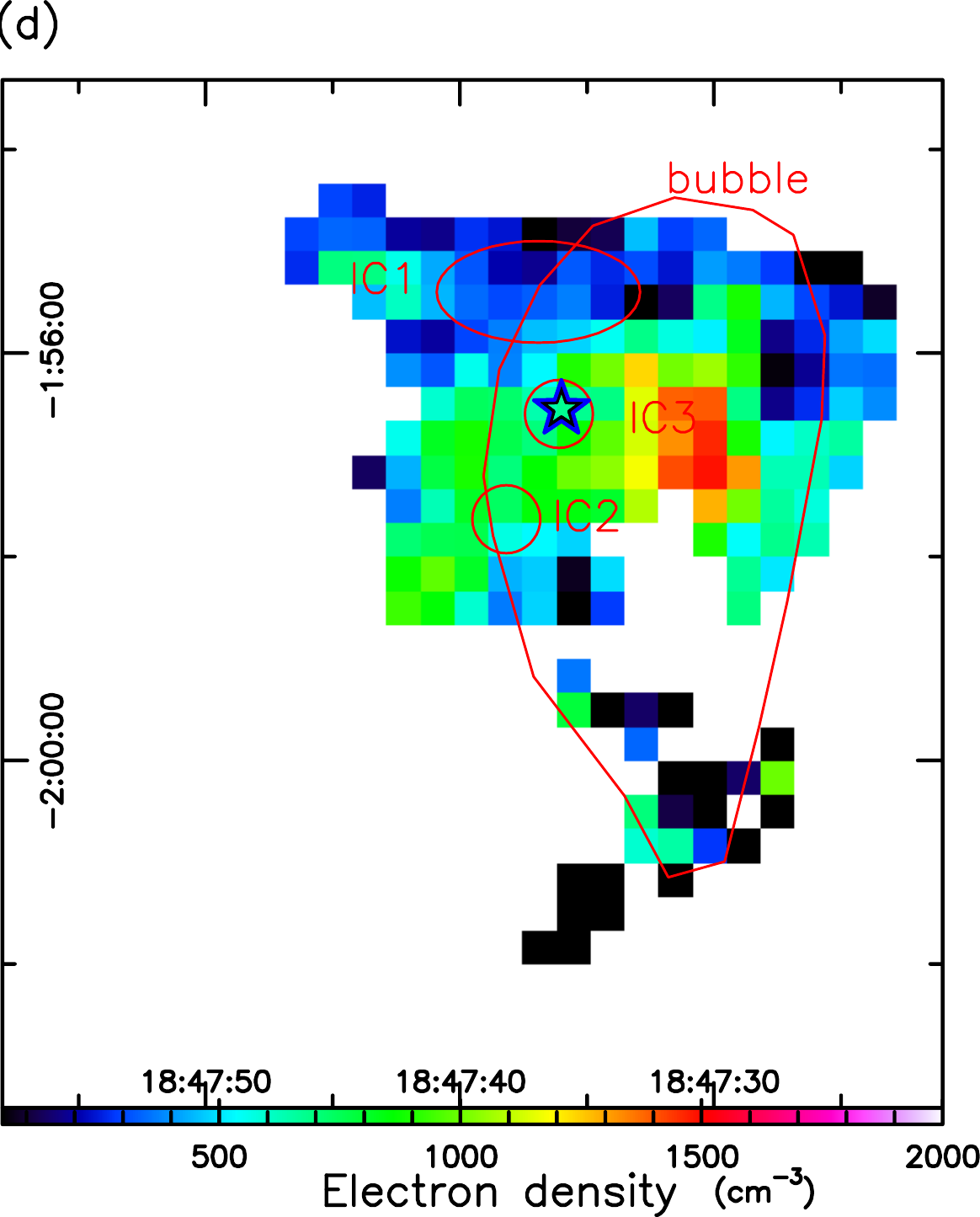}

\end{array}$
\caption{{\bf (a)}: The mm-RRL linewidths from Gaussian fits, 
 {\bf (b)}: The cm-RRL linewidths resulted from Gaussian fits, 
{\bf (d)}: The dynamical widths, 
 {\bf (e)}: The pressure widths,
Red ellipses indicate the position of the ionized gas bubbles and ionized clumps IC1, IC2, and IC3. Blue star marks the position of the W43 SSC \citep{blum99}.}
\label{fig:broaden}
\end{figure*}

\begin{figure*}[tbhp!]
\centering
$\begin{array}{c}
\hspace{-0.8cm}
\includegraphics[angle=0,width=18cm]{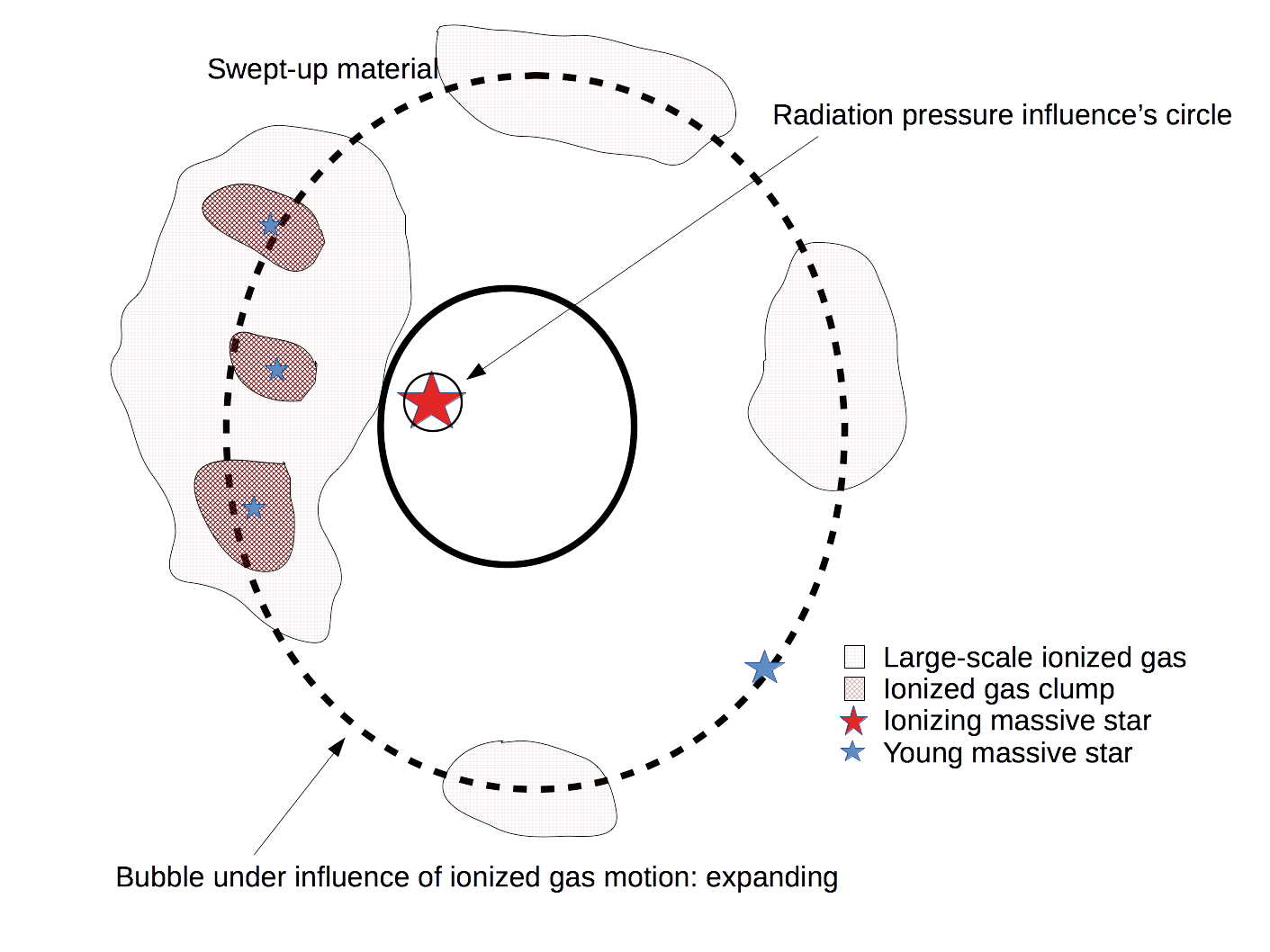} \\
\end{array}$
\caption{A sketch of the configuration of ionized gas in the vicinity of the W43-Main SSC. {\lda I don't think the "under the influence" statements are useful here} The sketch depicts the ionized gas bubble under the influence of gas dynamics, the cavity under influence of collisions with ions, and the circle of influence of radiation pressure.}
\label{fig:w43sketch}
\end{figure*}

Aside from tracing the bulk motion of the ionized gas using velocity shifts (see Figure\,\ref{fig:H92} and Section\,\ref{sect:expanding}), the mm-RRL and cm-RRL cubes enable us to map the intrinsic motions of the gas by probing different linewidth broadening distributions caused by macroscopic ionized gas motions (dynamic broadening) and impacts with ambient materials along their paths (collisional or pressure broadening). 
We determine the contribution of different broadening mechanisms to the measured line width using the procedure outlined by \citep{galvan-madrid12,keto08}. First, we calculate the FWHM maps pixel-by-pixel by fitting Gaussians to the mm-RRL and cm-RRL cubes. 
For this analysis, we smooth the mm-RRL cube to the 40\arcsec angular resolution and 10\,\kms spectral resolution of the cm-RRL cube, and regrid both mm-RRL and cm-RRL cubes to a $3\arcsec$ pixel grid. 
We fit only pixels that have an integrated intensity larger than 5$\sigma$, i.e 1.3 K km/s and 0.5 Jy/beam km/s (Figure\,\ref{fig:recombinationlines}a-b). 
The Gaussian fit to some representative spectra are plotted in Figure\,\ref{fig:spectra} and the resultant FWHM maps of cm-RRL (18--57\,\kms) and mm-RRL (20--44\,\kms) lines are illustrated in Figure\,\ref{fig:broaden}a-b.

While mm-RRLs' linewidths are mainly thermally broadened, $\Delta\nu_{\rm the}$, (due to microscopic gas motion, and dynamically broadened, $\Delta\nu_{\rm dyn}$, (due to macroscopic gas motion and turbulence), cm-RRL linewidths are further broadened by pressure from collisions with ions and electrons, $\Delta\nu_{\rm p}$.  For a pure-hydrogen gas with a given electron temperature $T_e$, which we assume to be $7030\pm50$\,K as observed in W43 using pointed single-dish observations \citep{quireza06}, the thermal broadening is given by:
\begin{equation}
\Delta\nu_{\rm the} = \left( 8{\rm ln}2k_{\rm B}\frac{T_{\rm e}}{m_{\rm H}}\right)^{1/2} = 17.94\pm0.03\,\kms
\label{eq:deltathermal}
\end{equation}
 where $k_{\rm B}$ is the Boltzmann constant, and ${m_{\rm H}}$ is the mass of hydrogen. 

Subsequently, $\Delta\nu_{\rm dyn}$ can be estimated from the $\Delta \nu_{\rm mm-RRL}$  as 
\begin{equation}
\Delta\nu_{\rm dyn}^{2} =  \sqrt{\Delta \nu_{\rm mm-RRL}^{2}-\Delta\nu_{\rm the}^{2}}
\end{equation}

 Conversely, $\Delta\nu_{\rm p}$ is solved from Equation 2 of \cite{keto08}, 
\begin{equation}
\Delta\nu_{\rm p} = 7.8{\Delta\nu_{\rm cm-RRL}}-\sqrt{46\Delta \nu_{\rm cm-RRL}^2 + 15\Delta \nu_{\rm mm-RRL}^2}.
\end{equation}

The resultant pixel-by-pixel maps of the dynamical and pressure broadening of the ionized gas W43-Main are plotted in Figure\,\ref{fig:broaden}c-d together with the mm-RRL and cm-RRL linewidth maps Figure\,\ref{fig:broaden}a-b. 

The spatial distributions of these two types of broadening are almost anti-correlated in 2-dimensional space.
The pressure broadening (4--55\,\kms) is dominated inside the bubble's cavity near the SSC and follows the cm-RRL linewidth distribution while the dynamical broadening (8--36\,\kms) is dominated in the bubble's edges and follows the mm-RRL linewidth distribution. 

The dominance of pressure broadening inside the bubble's cavity is further supported by the high electron density  $n_{\rm e}$  which we calculate as
\begin{equation}
{{n_{\rm e}}\over{10^5 \rm cm^{-3}}}    = \frac{1}{1.2}{{\Delta\nu_ p}\over{\Delta\nu_{\rm the}}} \left({{N}\over{92}}  \right)^{-7}\,, \\
\end{equation}
where N is the RRL principal quantum number. We obtain an electron density of $n_{\rm e} = 70--1500$\,cm$^{-3}$ for W43-Main, thus lie within the expected range for giant H{\scriptsize II}s region  \citep{quireza06}. 
The electron density decrease from the inner cavity to the outer envelope thus coincides with the distribution of pressure width. 

Dynamical broadening is dominant above the SSC and resembles the hot gas distribution as seen in the {\it Herschel} temperature map (Figure\,\ref{fig:recombinationlines}b)  which may suggest that large-scale dynamics of ionized gas are related with the hot neutral gas and dust, probably via the swept up gas activity.  
Further evidence supporting the hypothesis of the swept-up activity is the existence of the ionized gas clumps, IC1 to IC3, that are also low-density molecular clouds  hosting massive dense clumps \citep{motte03} and UCH{\scriptsize II} regions \cite{purcell13}. These clumps are located at a distance of 2 pc from the heating cluster. Similarly, on the other side of the bubble, there exists a candidate massive young stellar object \citep{saral17}, which might form under the influence of the large scale ionized gas flow which cause dynamical broadening of the RRLs.  

In the context of an H{\scriptsize II} region of a SSC, another source which could alter the dynamic of the ionized gas is the radiation pressure \citep{krumholz09b}. However, radiation pressure is only important in the inner part of the giant H{\scriptsize II} region driven by massive clusters whereas feedback from gas pressure have significant impact at larger radii. The characteristic radius at which gas pressure and radiation pressure \citep{krumholz09b} are equal is described as
\begin{equation}
r_{\rm ch} = 0.023S_{49} {\rm \, pc}
\end{equation}
where $S_{49}$, in units of 10$^{49}$ $s^{-1}$, is the ionizing luminosity of the ionizing source and $S_{49}=1$ is the Lyman$\alpha$ continuum luminosity of an average OB star. From a measured radio continuum integrated intensity $S_{\nu}=  55.55\pm0.22$ Jy \citep{bihr16} of the W43-Main ionized gas bubble from THOR continuum data, we can estimate the total number of Lyman ionizing photons $N_{\rm Ly\alpha}$ as described by \cite{mezger67} as:

\begin{equation}
\frac{N_{\rm Ly\alpha}}{8.9\times10^{46}\,\text{s}^{-1}} = \frac{S_{\nu}}{\text{Jy}}   \left(\frac{\phantom{N} \nu \phantom{N}}{\text{GHz}}\right)^{0.1} \left(\frac{T_{e}}{10^{4}\,\text{K}}\right)^{-0.45}   \left(\frac{d}{\text{kpc}}\right)^{2} \\
\end{equation}
where $\nu=1.436$\,GHz is the frequency of the radio continuum observations, $T_{\rm e}=7030$\,K is the electron temperature, and $d$=5.5\,kpc is the distance to W43-Main. We thus obtain a total Lyman$\alpha$ photon of $1.5\times10^{50}$ s$^{-1}$, and 
the characteristic radius at which the gas pressure and the radiation pressure equal is $\sim$0.4 pc. Therefore, radiation pressure is negligible on the scale of our consideration.

In summary, the dynamics of the giant H{\scriptsize II} region surrounding the W43-Main SSC is dominated by radiation pressure inside the radius of 0.4 pc from the cluster, by collisions with ions and electrons in the cavity zone about 2 pc from the cluster, and by large-scale ionized gas motion in the bubble zone at a distance larger than 2 pc. We sketch the configuration of ionized gas in W43-SSC in Figure\,\ref{w43sketch} for the purpose of visualization.




\section{Summary}
We show in this Letter the first map of large-scale emission of mm-RRL line emission toward the giant H{\scriptsize II} region around the W43 super star cluster. This high-spectral resolution mm-RRL cube shows direct evidence of an expanding ionized bubble that is interacting with the nearby dense neutral molecular dust ridge. Together with cm-RRLs, we make the first 2D map of electron density and of the different broadening mechanisms. We show that dynamical broadening, traced by the mm-RRL linewidth, is significant at the edge of the bubble where ionized clumps are located, whereas pressure broadening,  traced by the cm-RRL linewidth, are located inside the bubble's cavity where the electron density is higher. The large-scale mm-RRL maps show the potential of tracing the dynamics of the ionized gas, and is useful for studying star formation.  Follow up mm-RRL observations with ALMA will further reveal the dynamics of this region.

\begin{acknowledgements}
We thank Steve Mairs for improving the quality of the paper.
\end{acknowledgements}


\end{document}